\documentclass{emulateapj}

\shorttitle{Spectral Analysis using PCA}
\shortauthors{M. Ortiz and G. Galaz (2009)}

\begin{document}

\title{\textbf{S\lowercase{pectral} C\lowercase{lassification of} G\lowercase{alaxies using the} P\lowercase{rincipal}
  C\lowercase{omponent} A\lowercase{nalysis: a} W\lowercase{eb} B\lowercase{ased} T\lowercase{ool}}}

\author{Mauricio Ortiz and Gaspar Galaz}
\affil{Astronomy Department, Pontificia Universidad Cat\'olica de Chile,
    Santiago}

\begin{abstract}
We have developed a web tool to perform Principal
Component Analysis  (PCA, Murtagh \& Heck 1987; Kendall 1980) onto 
spectral data. The method is especially designed to perform spectral
classification of galaxies from a sample of input spectra, giving the set of
orthonormal vectors called Principal Components (PCs) and the
corresponding projections. The first two projections of the galaxy  
spectra onto the PCs are known to correlate with the morphological
type (Connolly et al. 1995) and, following Galaz \& de Lapparent (1998),
we use the parameters $\delta$ and $\theta$ which define a spectral
classification sequence of typical galaxies from ellipticals to late
spirals and star-forming galaxies. The program runs in the website
http://azul.astro.puc.cl/PCA/ and can be used without downloading any
binary files or building archives of any kind. 
\end{abstract}

\keywords{Spectral classification of galaxies: PCA ---
galaxies: spectral analysis}

\clearpage

\section{The Method: Brief Introduction}
In the context of Principal Component Analysis (PCA), each individual
galaxy spectrum is treated as a single data point in a
multi-dimensional space. The aim of the method is to find a set of
orthonormal vectors that best describe the intrinsic relation between
these data. These vectors are called Principal Components (PC's) and
satisfy the condition of minimal Euclidean distance from each data
point to the axes defined by this new base. A detailed description of
the PCA technique can be found in Murtagh \& Heck (1987), and in
Kendall (1980).\\ 
Connolly et al. (1995) have shown using the spectra from Kinney et
al. (1996), that the first 2 projections of the PCA define a sequence
tightly correlated with the morphological type. Following Galaz \&
de Lapparent (1998), two parameters are used for the spectral
classification of galaxies, due to its
geometrical meaning. These parameters are $\delta$ and $\theta$, and
are defined as follows:\\ 

\begin{equation}
 \delta = \arctan\left( \frac{\alpha_{2}}{\alpha_{1}}\right)  
\end{equation}
\begin{equation}
 \theta = \arcsin \left( \alpha_{3}\right) 
\end{equation}
\\
\\
\noindent where $\alpha_{1}$, $\alpha_{2}$ and $\alpha_{3}$ are the
projections of the corresponding spectrum onto the first, second and
third Principal Component, respectively. The physical meaning of
$\delta$ is the relative contribution of the red (or early) and the
blue (or late) stellar populations within a galaxy and $\theta$ is
related to the presence of significant emission lines in the spectrum.\\ 
It is worth mentioning that the definitions of $\delta$ and $\theta$
above, only makes sense if the first 3 projections carry most of the
contribution to the total variance, that is:
$\sqrt{\alpha_{1}^{2}+\alpha_{2}^{2}+\alpha_{3}^{2}}\sim 1$ (see Galaz \&
de Lapparent 1998).\\ 
\\
\\
\section{Program Input}
The program input must be a gzipped tar file (tar.gz). The file must
contain 1-D calibrated spectra (in the FITS 
format), both in flux and wavelength. It is assumed that the spectra
are completely reduced (sky and cosmic ray-subtracted, no fringing,
etc..) and corrected to the rest-frame wavelength. Each FITS spectrum must have a
unique name (no more than 50 characters) and a maximum number of
50.000 spectra is allowed by the code. It is possible that the spectra may have
different spectral coverage and resolution. The program will decide
the common wavelength range for all spectra and will reject each
spectrum whose initial and final wavelength differ by more than 3$\sigma$
from average. In the case of different resolution between spectra, the
program can rebin the spectra automatically (using a cubic spline
function) or by giving a proper wavelength step (in angstroms).\\ 

\section{Program Output}
The outputs of the analysis will be in a compressed tar file (tar.gz) and
can be downloaded after the ejecution of the program. Some important
results will be provided in the following files: 
\begin{itemize} 
\item DeltaTheta.txt: This file contains the values
  $\delta$ and $\theta$, calculated for each spectrum as defined
  above. 
\item Pc.txt: Here the user can find the Principal Components listed
  in descending order, being VEC-1 the first PC, VEC-2 the second PC,
  VEC-3 the third PC and so on. 
\item Projections.txt: In this file are the projections of each
  spectrum onto the eigenvectors (PC's), being PROJ-1 the projection
  onto the first PC, PROJ-2 the projection onto the second PC, PROJ-3
  the projection onto the third PC, and so on. 
\item Log.info.txt: This file stores some relevant information about
  the PCA concerning the spectra cutting and rebinning, as well as the
  spectral range covered and the resolution. 
\end{itemize}  

\noindent Inside the tar.gz file there is a README with detailed
information about each ouput. As a complementary result, after the
analysis it is possible to see some plots of the first PC's, and a series
of interactive plots of the first three projections and the
correlating parameters $\delta$ and $\theta$ , in which the user can
see each galaxy spectrum by clicking on the corresponding data point. 

\section{Special Considerations}

\begin{itemize}
\item For the calculation of the paremeters $\delta$ and $\theta$, it
  is assumed that
  $<\sqrt{\alpha_{1}^{2}+\alpha_{2}^{2}+\alpha_{3}^{2}} > \sim 1$ and
  only the first three projections are used, but depending on the
  quality of the spectra the user can look into the file
  Projections.txt to find the complete set of projections. 
\item The parameters $\delta$ and $\theta$ are calculated using the
  $\sqrt{\alpha_{1}^{2}+\alpha_{2}^{2}+\alpha_{3}^{2}} = 1$
  normalization as defined in Galaz \& de Lapparent (1998). 
\item The program can be used through the website
  http://azul.astro.puc.cl/PCA/. Please report problems, suggestions, and possible
  bugs to \textit{mortiz@astro.puc.cl} or
  \textit{ggalaz@astro.puc.cl}. 
\end{itemize}

\clearpage

\end{document}